\documentclass[preprint,journal]{vgtc} 

\usepackage{cite}
\usepackage[bottom]{footmisc}
\usepackage{mathptmx}
\usepackage{graphicx}
\usepackage{siunitx}
\usepackage{times}
\usepackage[T1]{fontenc}
\usepackage[latin1]{inputenc}
\usepackage{enumitem}
\usepackage[table]{xcolor} 
\usepackage{wrapfig}
\usepackage{multirow}
\usepackage{adjustbox}
\usepackage{mwe}
\usepackage{graphbox}
\usepackage{dirtytalk}
\usepackage{soul}
\usepackage{mdframed}
\usepackage{tabularx}
\usepackage{xcolor}
\usepackage{enumitem}
\usepackage{pdfpages}
\usepackage{dcolumn}
\usepackage[graphicx]{realboxes}
\usepackage{rotating}
\usepackage{graphicx}
\usepackage[capposition=bottom]{floatrow}
\usepackage{caption}
\definecolor{CindySalmon}{RGB}{232, 125, 114}
\definecolor{greenB}{RGB}{77, 175, 74}
\definecolor{purpleF}{RGB}{152,78,163}

\definecolor{highNegative}{HTML}{F2870C}
\definecolor{highPositive}{HTML}{22A2F2}
\definecolor{bothStrong}{HTML}{3EB82E}

\makeatletter
\AtBeginDocument{\let\hl\@firstofone}
\makeatother

\newcommand{\comment}[1]{}

\usepackage{balance}

\usepackage[bookmarks,backref=true,linkcolor=black]{hyperref} 
\hypersetup{
 pdfauthor = {},
 pdftitle = {},
 pdfsubject = {},
 pdfkeywords = {},
 colorlinks=true,
 linkcolor= black,
 citecolor= black,
 pageanchor=true,
 urlcolor = black,
 plainpages = false,
 linktocpage
}

\onlineid{1103} 
\vgtccategory{Theoretical and Empirical} 
\vgtcinsertpkg 

\title{How Do Viewers Synthesize \\Conflicting Information from Data Visualizations?}

\author{Prateek Mantri, Hariharan Subramonyam, Audrey L. Michal, and Cindy Xiong}
\authorfooter{ 
\item
Prateek Mantri, and Cindy Xiong are with the University of Massachusetts Amherst. 
E-mail: \{pmantri, cindy.xiong\}@cs.umass.edu. 
\item
Hariharan Subramonyam is with Stanford University. E-mail: harihars@stanford.edu.
 \item
Audrey L. Michal is with the University of Michigan Ann Arbor. E-mail: almichal@umich.edu. 
}

\shortauthortitle{Mantri \MakeLowercase{\textit{et al.}}: }

\abstract{Scientific knowledge develops through cumulative discoveries that build on, contradict, contextualize, or correct prior findings. Scientists and journalists often communicate these incremental findings to lay people through visualizations and text (e.g., the positive and negative effects of caffeine intake). Consequently, readers need to integrate diverse and contrasting evidence from multiple sources to form opinions or make decisions. However, the underlying mechanism for synthesizing information from multiple visualizations remains under-explored. To address this knowledge gap, we conducted a series of four experiments (N = 1166) in which participants synthesized empirical evidence from a pair of line charts presented sequentially. In Experiment 1, we administered a baseline condition with charts depicting no specific context where participants held no strong belief. To test for the generalizability, we introduced real-world scenarios to our visualizations in Experiment 2 and added accompanying text descriptions similar to online news articles or blog posts in Experiment 3. In all three experiments, we varied the relative direction and magnitude of line slopes within the chart pairs. We found that participants tended to weigh the positive slope more when the two charts depicted relationships in the opposite direction (e.g., one positive slope and one negative slope). Participants tended to weigh the less steep slope more when the two charts depicted relationships in the same direction (e.g., both positive). Through these experiments, we characterize participants' synthesis behaviors depending on the relationship between the information they viewed, contribute to theories describing underlying cognitive mechanisms in information synthesis, and describe design implications for data storytelling.
}


\begin{document}
\maketitle

\section{Introduction}
A goal of scientific research is to inform the public. 
To address the research-to-practice gap, scientists increasingly communicate key findings to a broader audience such as journalists, policymakers, and the general public to facilitate informed decision-making. 
Through easy-to-digest summaries and data visualizations, news articles make scientific knowledge accessible, engaging, and memorable~\cite{segel2010narrative, kosara2013storytelling, riche2018data}. 
However, news articles primarily focus on \emph{singular} discoveries or expert views, leaving readers to form opinions across multiple articles with potentially conflicting evidence~\cite{carreyrou2019bad, brand2011solving}, especially for controversial topics~\cite{van2014dealing}. 
For example, throughout the COVID-19 pandemic, numerous pieces of conflicting information have emerged (e.g., regarding the effectiveness of masks and treatments), and a large portion of the public is exposed to conflicting evidence~\cite{nagler2020public}. 
Even a comprehensive search might not point to one simple, unambiguous conclusion~\cite{kobayashi2009influence}, leaving the responsibility of synthesizing information to draw an actionable conclusion to the consumer. 

For non-experts, synthesizing information is not an easy task. 
Even trained scientists are still figuring out how to effectively synthesize high volumes of information. 
Due to the incremental nature of scientific research, it is common for subsequent work to offer contradictory results~\cite{ohlsson1994systematic}. 
Scientists often write reviews of existing work by carefully assessing evidence, making comparisons, and building connections across studies~\cite{cook1997systematic}. 
To address bias in interpretation, researchers have called for systematic reviewing guidelines and meta-analysis to enforce statistical procedures for synthesizing findings~\cite{tranfield2003towards, hedges1992meta}. 
However, general audiences reading articles online are likely not equipped with the resources to conduct statistically rigorous meta-analyses. 
Consequently, they might rely on heuristics to filter sources to answer their questions and be more susceptible to cognitive biases when synthesizing information~\cite{azzopardi2021cognitive, kahneman2011thinking}.

For example, during the information-seeking stage, people may fall victim to the satisfaction of search bias (a.k.a. premature closure), where they stop looking for additional information once a plausible result is found~\cite{norman2017causes, fleck2010generalized}. 
In addition, people may exhibit confirmation bias and overly weigh the information that agrees with their pre-existing beliefs while discounting information against it~\cite{kahan2017motivated, nickerson1998confirmation}. 
Additionally, people may be impacted by Bandwagon effects, taking on a similar point of view as the majority of voices they hear~\cite{kelly2010effects}. 
These effects are amplified with search engines as information seekers are more likely to go with the suggested query~\cite{harris2019detecting}. 
Furthermore, these biases generalize across both search for facts and search for exploration~\cite{kim2008task, palmquist2000cognitive, white2001questions}. 

While existing work has surveyed biases that occur during information seeking~\cite{azzopardi2021cognitive}, we know very little about how people informally synthesize information. 
Our primary motivation for this research is to understand the underlying cognitive mechanisms of non-experts when synthesizing multiple sources of scientific evidence. 
This understanding will enable researchers to create tools for non-experts to synthesize information effectively and make less biased data-driven decisions.
\newline
\vspace{-2mm}

\noindent \textbf{Contributions:} We contribute a preliminary model of \textit{how people synthesize conflicting visual information} to inform the design of search tools, learning tools, and visual analytic tools. 
In three empirical studies, we presented participants with conflicting scientific information comprised of visualizations and text. 
We asked participants to synthesize the information and report their mental representation of the relationship between the depicted variables. 
Using this data, we model how readers synthesize conflicting evidence and identify systematic biases in information synthesis. 
We additionally contribute a survey of people's current practices in seeking information, synthesizing information, and resolving conflicting information online.

\section{Related Work}
\subsection{Information Synthesis and Decision Making}
Information synthesis involves foraging and sensemaking cycles~\cite{pirolli2005sensemaking}. 
Foraging is often driven by readers' need for cognition~\cite{ leman2013beliefs}. 
There is a large body of work on what factors motivate the search for information, but that is outside the focus of our work. 
Rather, we are interested in how non-experts synthesize evidence~\cite{kuhlthau1991inside, morris2010comparison}. 
Although scientists have developed rigorous techniques for synthesizing information from multiple sources, including aggregative, integrative, interpretive, and explanatory synthesis ~\cite{rouse2017interactive, briner2012systematic}, non-expert readers typically lack these skills and rely on their intuitive sensemaking abilities~\cite{kapon2017unpacking}. 
Broadly, non-experts tend to rely on System 1-type heuristics when evaluating the quality or reliability of scientific evidence; for example, people's evaluations of scientific evidence are often influenced by contextual features such as anecdotes, formulas,  and brain images~\cite{shah2017makes}. 
In the context of data visualizations, although some statistical features can be extracted quickly by the visual system (such as outliers or the overall mean), computing relations between specific data points or trends requires slow sequential processing~\cite{franconeri2021science}. 
Because evidence synthesis relies on careful comparison and appropriate weighting of evidence, it is unlikely that people can synthesize multiple visualized data sets (such as dashboard displays) quickly and accurately~\cite{few2009now}. 
Thus, it is crucial to establish how accurately people can visually synthesize various combinations of visualizations. 
Given the broader challenges of organizing, tracking, evaluating, and synthesizing evidence online, decision-making tools may be necessary to support the evidence synthesis process among non-experts~\cite{liu2019unakite,subramonyam2020texsketch}. For example, Unakite~\cite{liu2019unakite} allows users to drag snippets of information found online to a comparison table that keeps track of findings and considers tradeoffs of specific features when creating new software. A similar comparison table approach could help users synthesize heterogeneous evidence from multiple sources online. However, we first need to know more about how users perceive and interpret conflicting evidence from visualizations to inform the design of such decision-making tools.
\vspace{-0.5mm}
\subsection{Processing Conflicting Information}
Prior research has investigated conflicting information as a \textit{selection} problem, uncovering numerous biases in how people choose between information sources. 
\textit{Confirmation bias} is widely prevalent, in which readers favor confirmation-seeking information over contradicting information~\cite{klayman1995varieties, charness2021people, knobloch2020confirmation, knobloch2012preelection}. 
This behavior also applies to chart comprehension; readers selectively attend to information aligned with their beliefs~\cite{ellis2018cognitive}. 
For example, Democrats and Republicans differed in their interpretations of mortality rates when the same chart was labeled as COVID-19 or Influenza~\cite{ericson2022political}. 
Readers tend to impose categorical distinctions (i.e., \textit{binary bias}) when interpreting visualizations of continuous data, resulting in distorted beliefs ~\cite{fisher2018binary}. 
Furthermore, news sources make use of deceptive tactics to amplify or nudge biased information seeking~\cite{garz2020supply}. 
Readers misinterpret charts when presented with an inverted or truncated axis~\cite{lauer2020people}. Frames and slants in visualization titles lead viewers to perceive opposing information for the same visualization~\cite{kong2018frames}. Slanted information also amplifies \textit{congeniality bias}, and readers favor belief validation over information accuracy~\cite{hart2009feeling}.  

However, few studies have examined conflicting visual evidence as a \textit{synthesis} problem. Prior work studying how people process text with conflicting evidence has proposed models for processing belief-inconsistent information through strategic \textit{elaboration} of inconsistencies~\cite{richter2017comprehension}. Nevertheless, initial attitudes and beliefs can lead to elaboration favoring prior stances. For instance, when students wrote an essay based on conflicting sources, those with a strong initial stance included a large proportion of information not presented in the text ~\cite{van2014dealing}. In contrast, students with a neutral attitude included more information from the materials. 

In some cases, readers may be intrinsically motivated to resolve conflicting information through information seeking and synthesis (i.e., uncertainty can be a driver for information acquisition and processing)~\cite{ford2004modeling}. However, recent work has shown that when readers are presented with belief-incongruent correlational visualizations, uncertainty visualizations only lead to minor belief updating compared to visualizations without uncertainty~\cite{karduni2020bayesian}. Additionally, non-scientists do not always perceive conflicting findings as scientific progress and can interpret conflicting findings as a loss of scientific knowledge ~\cite{koehler2019public}. This paper aims to address gaps in understanding what strategies readers apply when presented with conflicting visualizations.  

\vspace{-1mm}
\subsection{Visualization Perception and Memory}

Synthesizing information relies heavily on the recall of previously viewed information; however, our memory for information can be biased in many ways. For instance, primacy and recency effects, where the first and last items in a sequence are better remembered \cite{li2010primacy}, are well documented \cite{altmann2000memory, cowan2002deconfounding}. 
Similar to order effects, high magnitude values are weighted more heavily when people must summarize multiple values presented over time \cite{tsetsos2012salience}. 
The salience of these events (first or peak event; high magnitude value) explains their over-weighting.
People are also sensitive to categorical boundaries when retrieving information \cite{mccoleman2020no, parkes2001compulsory}. 
For example, categorical boundaries between hues can exaggerate differences between those that straddle a boundary compared to hues that do not \cite{bornstein1984discrimination}.
When making numerical estimations, people are systematically biased to overestimate small numbers and underestimate large numbers \cite{xiong2022investigating, ceja2020truth, mccoleman2020no}.
Furthermore, the number of data points can influence viewer perception \cite{cowan2001magical, hollingworth2008understanding, zhang2008discrete}. 
For example, adding more data points to a visualization can increase the difficulty of recalling data values \cite{mccoleman2021rethinking}. 
Beliefs can also distort information extraction \cite{xiong2022seeing} and memory \cite{smeets2009s}; high plausibility and firm belief that an event has occurred can drive people to recall false memories \cite{scoboria2006effects}.

Information context can also influence memory in visualizations.
For instance, participants are more accurate at recalling average positions from bar charts compared to line charts \cite{xiong2019biased}.
The aspect ratio also plays a part, such that people recall bars with an aspect ratio closer to a 1:1 square more accurately 
\cite{ceja2020truth}. 
When data are integrated with illustrations, people remember it better than minimalistic charts \cite{bateman2010useful, borgo2012empirical}.
Placing symbolic numbers on or near visual data encoding marks to generate ``data redundancy'' can also improve memorability \cite{borkin2015beyond}.
Highlighting and annotation improve memorability, such that data visualizations that are decluttered and focused are more accurately recalled 
\cite{ajani2021declutter, subramonyam2018smartcues}.
However, studies show that verbal labels of data can interfere with memory accuracy \cite{droit2008time}.
For example, the wording of a visualization title can bias people to remember the strength of the depicted relationship differently \cite{kong2018frames}. 
Our study accounts for these potential memory distorting effects through carefully designed experiments.

\begin{figure*}[h!]
\centering
  \includegraphics[width = \linewidth]{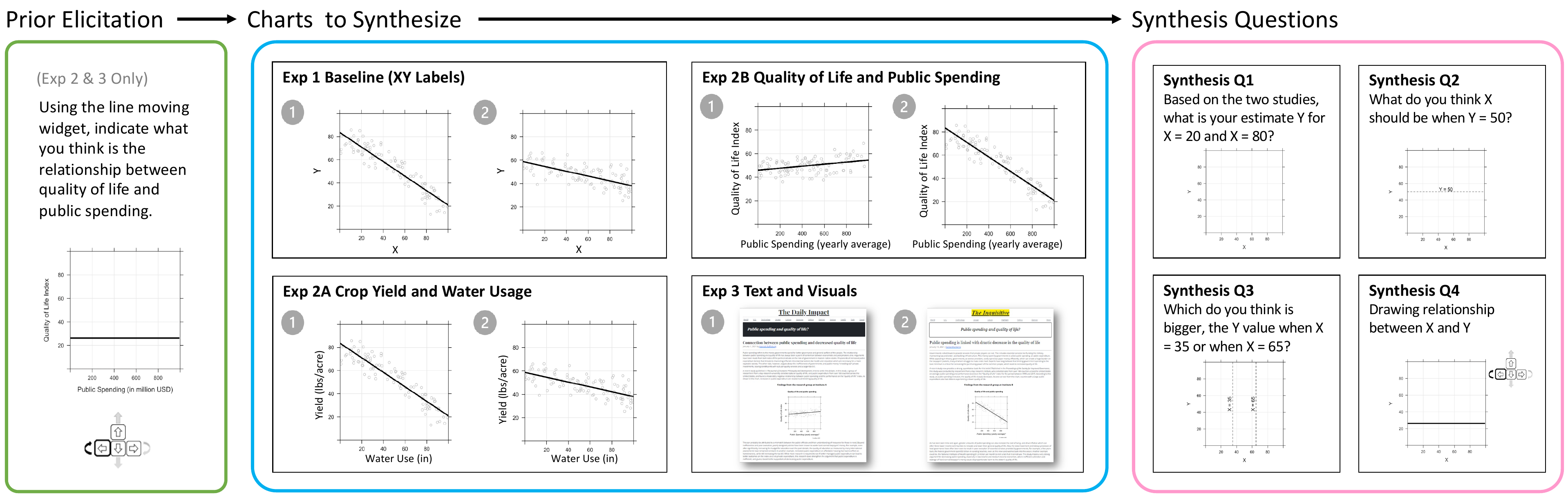}
  \caption{A schematic of the experiment flow, including example stimuli and survey questions.}
    \label{fig:schematicExperiment}
\end{figure*}

\section{Study Motivation and Overview}
\label{StudyOverview}

Informed by prior work, we investigated characteristics of scientific communication, including the topic, the thesis of the argument, and the data supporting the argument \cite{pandey2014persuasive}, with a focus on how people synthesize multiple pieces of evidence in the form of visualized data. 
In \textit{Experiment 1}, we showed data devoid of any contextual information to establish a baseline for how visual information synthesis occurs perceptually. 
In \textit{Experiment 2}, we provided participants with contextual information via two real-world scenarios devoid of any argument or interpretation of the data. 
In \textit{Experiment 3}, we provided all three necessary information points - context, argument, and data to support the argument  - to create a comprehensive real-world scenario. 
Participants viewed two news articles with text descriptions that included interpretations of the data. 
As information pertinence influences motivation \cite{pandey2014persuasive, luo2019motivated}, we chose a scenario of potential relevance to participants - the relationship between quality of life and public spending. 
Because visualizations are often studied independently of their context, despite appearing with text in real-world scenarios (e.g., \cite{ajani2021declutter, borkin2013makes}, Experiment 3 helps us understand the impact of text accompanying visualizations.
This approach better simulates real-world scenarios and increases the ecological validity of the work. 
Our experimental stimuli, analysis, and data can be found in the supplementary material (hereon referred to as SM) at~\url{https://osf.io/pc3f7/}.

\comment{
\begin{figure}[t]
\setlength{\textfloatsep}{4pt plus 1.0pt minus 1.0pt}
\setlength{\intextsep}{4pt plus 1.0pt minus 1.0pt}
\setlength{\floatsep}{4pt plus 1.0pt minus 1.0pt}
\centering
  \includegraphics[width=\textwidth]{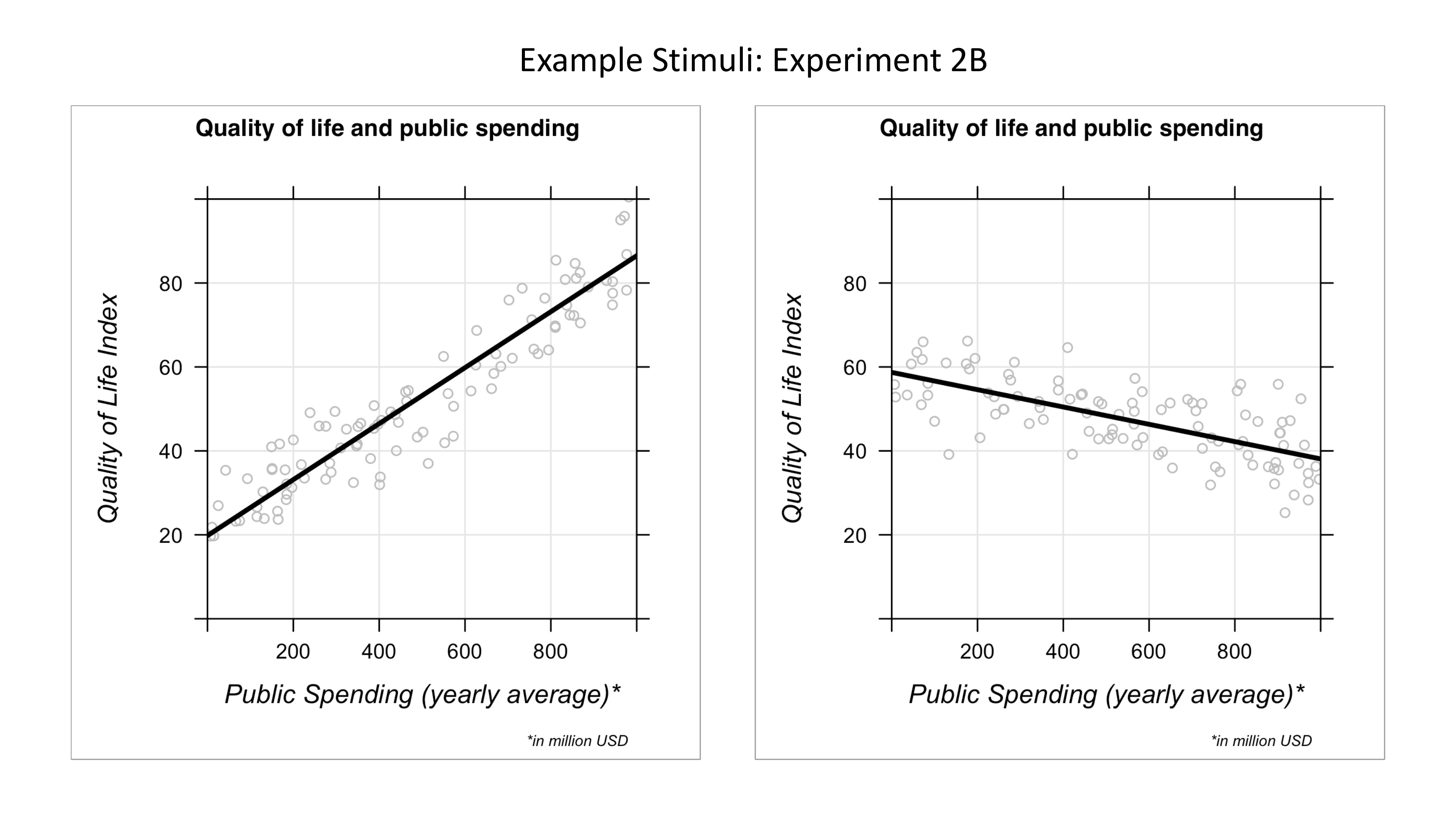}
  \caption{A sample from the six charts we used as stimuli for Experiments 2B, and 3. The same charts were used across all 4 Experiments, with minor changes in axes labels, and scales to reflect realism in the metrics displayed through them.}
    \label{fig:exampleStimuli}
\end{figure}
}

\begin{figure*}[t!]
\setlength{\textfloatsep}{4pt plus 1.0pt minus 1.0pt}
\setlength{\intextsep}{4pt plus 1.0pt minus 1.0pt}
\setlength{\floatsep}{4pt plus 1.0pt minus 1.0pt}
\setlength{\dbltextfloatsep}{6pt plus 1.0pt minus 2.0pt}
\setlength{\dblfloatsep}{6pt plus 1.0pt minus 2.0pt}
    \includegraphics[width=\linewidth]{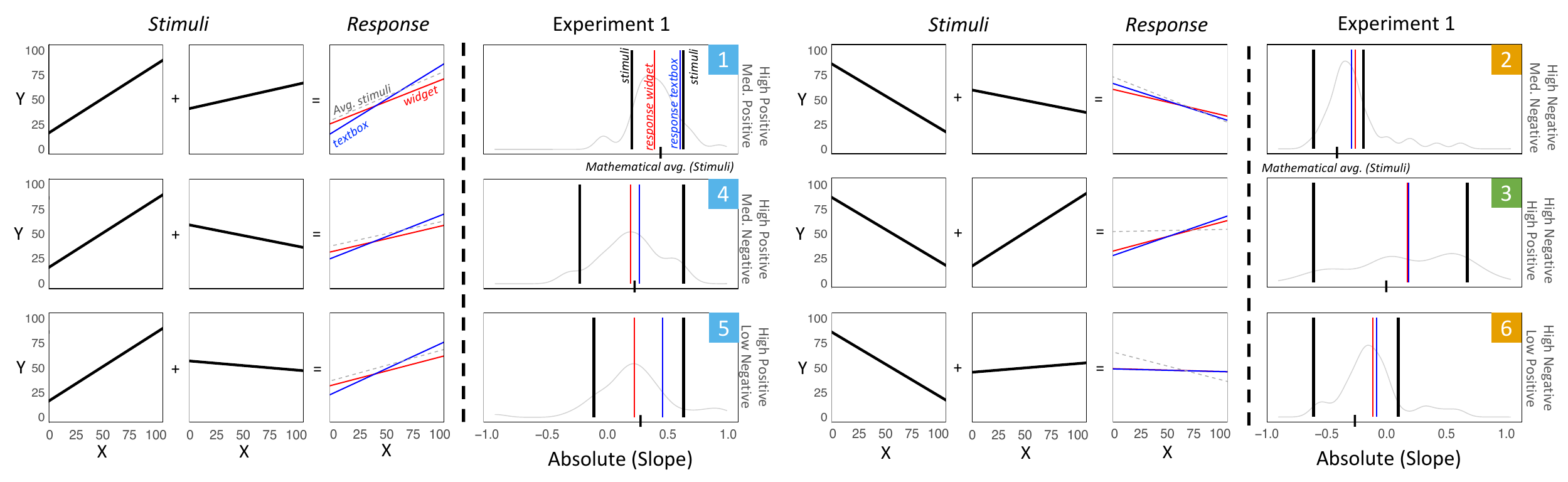}
         \caption{The average participants' response lines as computed through the custom-made widget and responses on text boxes. On both left and right panels, the first two columns denote the presented stimuli, and the responses have been displayed in column 3.  Column 4 displays the position of stimuli and response slopes on an absolute slope scale. The density plot shows the distribution of the participant responses. 
         The number on the top right corner in column 4 denotes the condition number. As an example showing data for Experiment 1, the left panel shows the conditions that contain a \textcolor{highPositive}{High positive} slope, while the right panel shows the conditions that contain a \textcolor{highNegative}{High negative} slope. 
         The color of the condition number box denotes the presence of either \textcolor{highPositive}{High positive} (blue), \textcolor{highNegative}{High negative} (orange), or \textcolor{bothStrong}{both} (green). Note: \hl{For computing the average participants' response, the x-axis values for Experiments 2B, and 3 were rescaled by a factor of 0.1}. Analyses of other Experiments can be found in the SM.}
             \label{fig:mathematicalAverage_exp1}
\end{figure*} 

\section{Experiment 1 --- Establishing baseline}

We presented participants with two visualizations, as shown in Figure \ref{fig:schematicExperiment} but without any context. Then, participants were asked to synthesize the information to estimate the relationship between the variables. 

\subsection{Participants}

We recruited 300 participants using \hl{Prolific.co}~\cite{palan2018prolific}. Based on pilot data ($N=78$), we conducted a power analysis that suggested a target sample of 288 participants would yield 80\% power to detect a signal at a significance level of 0.05. Considering potential data losses, we chose a sample size of 300. We filtered for participants living in the United States and fluent in English and excluded participants who had failed attention checks or provided nonsensical responses.
We ended up with 295 participants ($\mu_{age} = 34.9$, $\sigma_{age} = 12.74$, \text{139 women}). \hl{Participants were compensated at a flat rate of \$2.54 with an expected time commitment of 13 minutes, consistent with the average hourly wage of about \$9.50 at Prolific.co. However, since our time expectations were generous, participants were practically compensated at \$11.22 per hour for their effort.} The participant pool had a good variance in education levels. See the SM for a summary of participant demographics. To reduce the risk of variance in appearance and relative size of the stimuli, we asked participants to adjust the on-screen content size using credit card dimensions prior to the start of the Experiment. The entire Experiment was created in jsPsych~\cite{de2015jspsych}, and all data was stored on cognition.run servers~\cite{congintionrun}.

\vspace{-0.5mm}
\subsection{Materials and Design}
We created six line charts with varying \textit{slopes} and \textit{intercepts} to depict six strengths of correlation between two variables - (a) High negative (b) Medium negative (c) Low negative (d) Low positive (e) Medium positive (f) High positive. For this Experiment, the scales for both x and  y axes varied from 0 to 100 (see Figure \ref{fig:schematicExperiment}). 

Each participant was randomly assigned to one of six Conditions: (1) \textcolor{highPositive}{High positive and Medium positive slope} (2) \textcolor{highNegative}{High negative and Medium negative slope} (3) \textcolor{bothStrong}{High negative and High positive slope} (4) \textcolor{highPositive}{High positive and Medium negative slope} (5) \textcolor{highPositive}{High positive and Low negative slope} (6) \textcolor{highNegative}{High negative and Low positive} slope. Across all Conditions, the stimuli were counterbalanced for order of presentation.

\vspace{-0.5mm}
\subsection{Procedure}
Participants started the Experiment with informed consent, followed by a screen resizing widget to adjust the resolution (\hl{refer to Figure} \ref{fig:schematicExperiment} \hl{for a general schematic of the three Experiments}). Next, participants read a hypothetical scenario describing the two charts as research outputs regarding two unknown variables, $X$ and $Y$, from two equally reputable institutions. They then viewed the first chart and, on a new page, the second chart. On these pages, we instructed participants to pay close attention to the charts to memorize the relationship between the variables. After each stimulus, participants reported the abscissa for a given ordinate to ensure they attended to the chart information. 


Once participants had seen both charts, we captured their mental model of the relationship between the two variables via a custom-built widget (\hl{Synthesis Q4 in Figure} \ref{fig:schematicExperiment}).
Participants were prompted to synthesize data from the two charts by adjusting the slope and intercept of a line in an empty chart identical in size and style to the previous charts.
They then explained how they synthesized the charts via a free response question.
Their synthesized response was triangulated using two additional questions presented right after the stimuli sequence: \hl{one question asked for the values of the variable $Y$ for the corresponding values of the variable $X$, and one question asked for the value of $X$, for a particular $Y$ (Synthesis Q1 and Q2 in Figure} \ref{fig:schematicExperiment}). \hl{Participants were also given a decision-making task for maximizing the probability of getting a particular value of $Y$ for two given values of $X$}. 
As shown in Figure \ref{fig:schematicExperiment}, for all three questions, participants were provided with graphical aids to help them remember the information presented to them earlier (\hl{see SM for a detailed view of the Experiments}).

Since this Experiment required participants to be conversant in visual information, we measured their graph literacy. We used the abridged version (5 items) of the Subjective Graph Literacy (SGL) tool \cite{garcia2016measuring} for its robust psychometric properties. 
We also measured participants' attitudes towards science and scientific research by adapting relevant items from existing surveys \cite{funk2019trust}\cite{nadelson2014just} to control for any effect arising from prior beliefs and attitudes on scientific research. We also included an attention check and one reverse-coded item as a sanity check. Participants' responses are reported in the SM.

Towards the end of the survey, participants were asked free-response questions about their strategies for dealing with conflicting information on the internet, the relatability of the scenarios presented, and basic demographic information. We also measured how much effort it took participants to complete this task. Results from these qualitative questions are discussed in Section \ref{Exp4QualData}.

\vspace{-0.5mm}
\subsection{Results} 
We captured participants' information synthesis using a custom-made widget. After viewing the stimuli, participants were asked to report their estimation of the relationship between the two variables of interest. Their responses can be seen in Figure \ref{fig:mathematicalAverage_exp1}. 

As shown in Figure \ref{fig:mathematicalAverage_exp1}, participants diverged from the mathematical average of the two stimuli slopes, except for Conditions \textcolor{highPositive}{1} and \textcolor{highPositive}{4} (High positive/Medium positive, High positive/Medium negative). The mathematical average here denotes the arithmetic mean of the two stimuli slopes. For Conditions \textcolor{highNegative}{2}, \textcolor{bothStrong}{3}, and \textcolor{highNegative}{6}, involving \textcolor{highNegative}{a High negative stimulus}, we found a statistically significant divergence from the mathematical average with small to moderate effects (see SM for details). Also, for Conditions with \textcolor{highPositive}{one High positive stimulus}, the difference between the stimuli average and participants' average response was minimal. This suggests that participants were more open to the plausibility of a High positive correlation between the two variables than a negative relationship. 


\vspace{-2mm}
\subsubsection{Triangulation} 
Researchers in psychometrics and learning sciences have demonstrated that the way a question is asked may affect responses \cite{breuer2020effects}\cite{goldin2015framing}. 
When determining participants' mental models for the synthesized data, we wanted to ensure that our measurement was both reliable and valid. For measuring synthesis, as described in the \textbf{Procedure} section, we asked three more questions beyond the drawing task to triangulate our measurement of participants' estimation of the relationship between the depicted variables. This section covers the agreement statistics between different types of questions and tasks we gave our participants to ascertain their mental models.

We followed three approaches for triangulating, first by retrieving the values from the line-drawing widget \hl{(Synthesis Q4 on Figure} \ref{fig:schematicExperiment}) at the specific points ($x=20$ and $80$, and  $y=50$) that we also asked for via the text input from  participants \hl{(Synthesis Q1 and Q2)}.
We then compared the widget and the text input values to see if there were statistically significant differences. A paired sample t-test on the three different coordinates ($x = 20$ and $80$, and  $y = 50$) revealed statistically significant differences for $x=80$ ($p=0.045$), but not for $x=20$ ($p = 0.073$), or $y=50$ ($p=0.818$). To test whether these differences were localized within some Conditions, we ran 12 independent t-tests for \hl{each of the three values: $x = 20$, $x = 80$, and $y = 50$. We also ran six other t-tests for each of these values, accounting for order effects} (see SM). Differences were localized to two Conditions - \textcolor{highPositive}{High Positive/Medium positive} and \textcolor{highPositive}{High positive/Low negative}.

Further, we computed the slope from the values that participants inputted using text boxes \hl{(Synthesis Q1)} and compared it against the slope value drawn using the widget \hl{(Synthesis Q4)}. For 3 out of the 12 Conditions, the slopes drawn using the widget and those computed using the abscissa values were significantly different ($p < 0.05$, see SM for details). These analyses suggest that participants' mental models might depend upon the modality of the question asked. Further research is required to understand people's information retrieval strategies while encountering graphical information.

Lastly, as a third mode for triangulating visual information synthesis, we asked a decision-making question to optimize the ordinate for two abscissae choices \hl{(Synthesis Q3)}. We evaluated what the decision choice would be under 4 scenarios: 1) if a mathematical average were taken for each of the six Conditions, 2) based on the input on the widget\hl{(Synthesis Q4)}, 3) if a slope were computed based on the values collected via text-box responses\hl{(Synthesis Q1)}, and 4) participants' actual decision choice \hl{i.e., response on Synthesis Q3}. Based on this analysis, all categories had over $60\%$ agreement, with the highest agreement ($96\%$) between widget and text boxes \hl{(Synthesis Q1 and 4)}. All other categories had $61$-$77\%$ of agreement levels. This shows that the different modes of capturing mental models generally function equally well, with a stronger relationship between data captured through text box and widget. 

\vspace{-0.5mm}
\subsubsection{Order}
We computed the normalized distance from the stronger slope and used linear regression to see whether the order in which the stronger slope was presented influenced this distance metric.(DV: Normalized distance from stronger slope, IV: Order of the second slope $R^2_{adj} = -0.003$, $p-value=0.746$). However, we did not find any order effects for Experiment 1. 

\vspace{-0.8mm}
\subsubsection{Direction of the presented stimuli}
\begin{figure*}[t!]
\centering
  \includegraphics[width=\columnwidth]{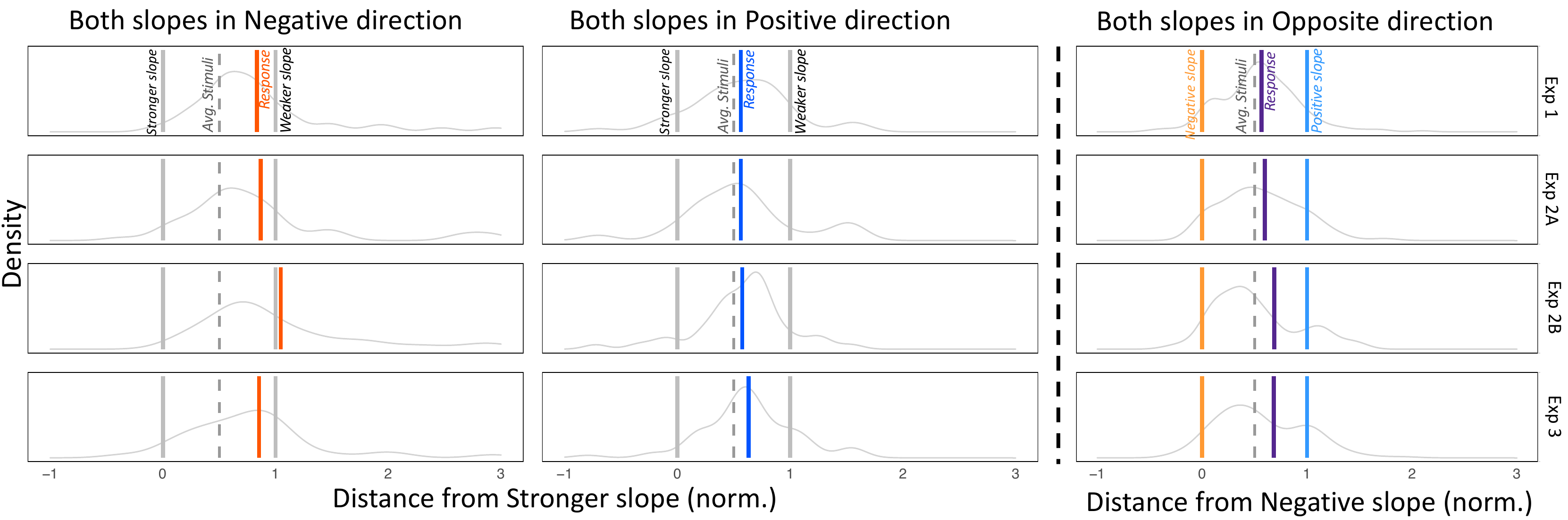}
  \caption{Effect of the \textit{direction} of stimuli slopes on two distance metrics. Columns 1 and 2 depict the effect of having stimuli slopes in the same directions on the normalized distance of the response from the stronger slope. Column 3 depicts the effect of having stimuli slopes in opposing directions, and how far the response slope moves from the negative slope on a normalized distance metric}
    \label{fig:effectDirection}
\end{figure*}

\hl{We measured the distance between the synthesized slope and the two stimuli slopes after normalizing the distance between the stimuli slopes on a scale of 0 to 1, with the stronger slope being 0. The distance between the synthesized and stimuli slopes on this normalized scale has been called weight. This metric shows participants' favoring of one stimulus slope over the other. }As shown in Figure \ref{fig:effectDirection}, when the two slopes were in the \textit{same} direction (both positive or both negative), the synthesized slope weighted the stronger slope $40.85\%$ and the weaker slope $59.15\%$ ($SE = 2.5\%$), accounting for order effects, when analyzed together. However, \hl{these differences in weighting became even more pronounced} when both slopes were negative, with \hl{significantly} more weight assigned to the weaker slope - 83.34\% ($SE = 9.26\%$)
When the two slopes were in the \textit{opposite} direction, the synthesized slope weighted the positive slope $56.49\%$ ($SE = 2.61\%$), accounting for order effects.

\vspace{-0.8mm}
\subsubsection{Discrepancy between the presented stimuli}
\begin{figure*}[t!]
\setlength{\textfloatsep}{4pt plus 1.0pt minus 1.0pt}
\setlength{\intextsep}{4pt plus 1.0pt minus 1.0pt}
\setlength{\floatsep}{4pt plus 1.0pt minus 1.0pt}
\setlength{\dbltextfloatsep}{6pt plus 1.0pt minus 2.0pt}
\setlength{\dblfloatsep}{6pt plus 1.0pt minus 2.0pt}
\centering
  \includegraphics[width=\linewidth]{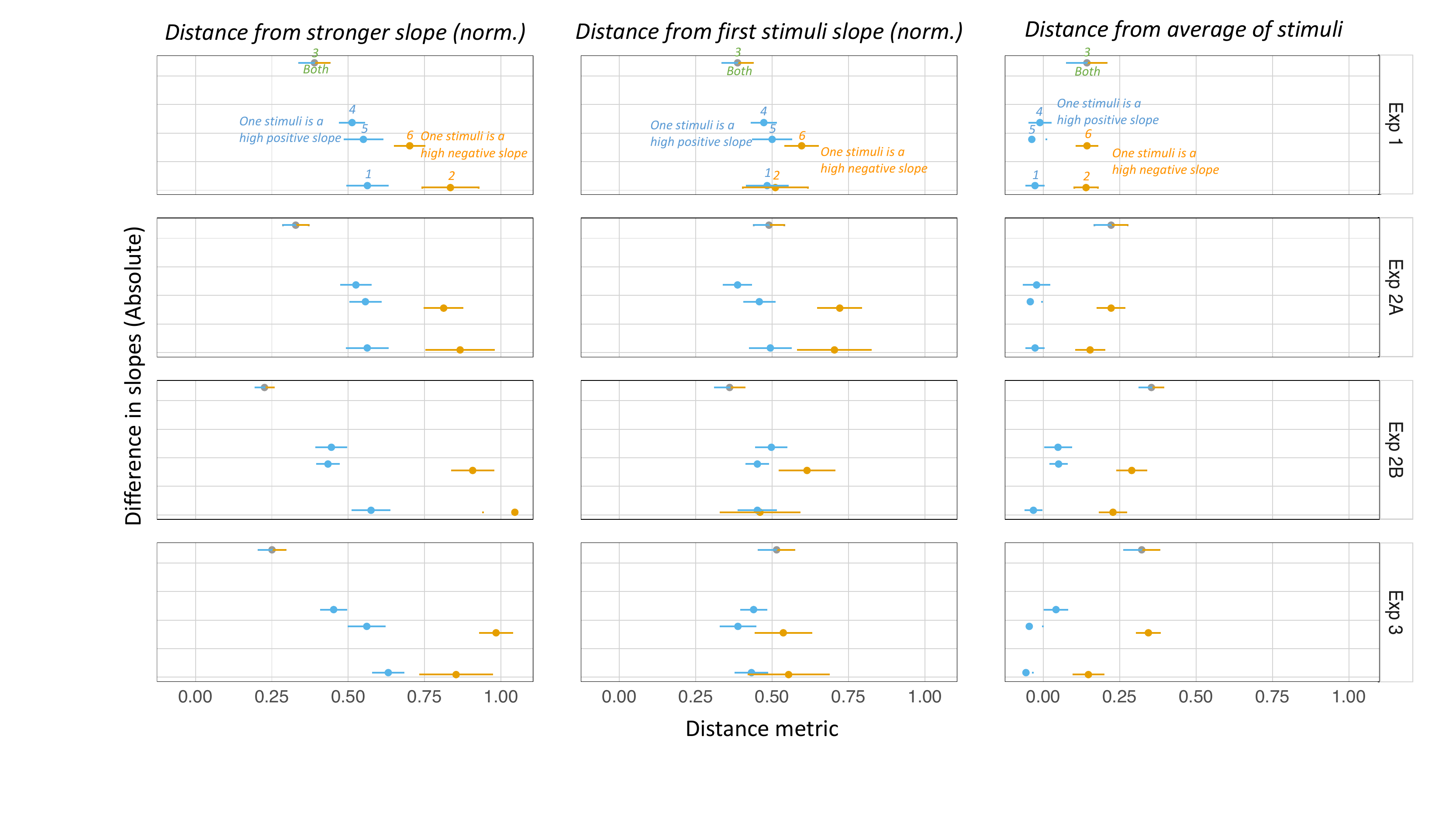}
  \caption{
  We computed three distance metrics to understand how divergent information (i.e., the varying slopes of the stimuli) affects visual synthesis.
  \textbf{Left:} The difference between participants' synthesized slope and the stronger slope (among the two line charts they viewed) is negatively correlated with the difference between the two stimuli slopes. The more discrepancy between the stimuli slopes, the more similar the synthesized slope is to the stronger slope. 
  \textbf{Middle:} The difference between the synthesized slope and the first slope seen is negatively correlated with the difference between the two line chart slopes. The more discrepancy between the stimuli slopes, the more participants gravitate towards the first slope when synthesizing.
  \textbf{Right:} The difference between participants' synthesized slope and the mathematical average of the two stimuli slopes is positively correlated with the difference between the two stimuli slopes. The more discrepancy between the stimuli slopes, the further participants' synthesized slope is from their mathematical average.
  }
    \label{fig:errorBars}
\end{figure*}

We analyzed participants' perceptions of the relationship and difference between the presented stimuli slopes using three different distance metrics: distance from the first presented stimulus, distance from the stronger stimulus, and distance from the mathematical average. We observed a \textcolor{highNegative}{High negative} - \textcolor{highPositive}{High positive} divide across all three distance metrics (see Figure \ref{fig:errorBars}). \hl{Conditions with a} \textcolor{highPositive}{High positive} \hl{slope seemed to always be on the left side of the Conditions that included a} \textcolor{highNegative}{High negative} \hl{slope.} Furthermore, as the difference between stimuli slopes increased, response slopes tended to move away from the mathematical average. In contrast, as the difference between stimuli slopes increased, response slopes tended to move closer to the stronger slope.

\section{Experiment 2 --- Real world scenarios}

We investigated how participants synthesize conflicting information from real-world scenarios involving charts published by top research institutes. We provided participants with two different contexts, including (a) yield in a hypothetical crop as a function of water use and (b) quality of life as a function of public spending. Similar to Experiment 1, participants were expected to synthesize this information to generate a mental model of the relationship between the two variables.

\subsection{Participants}
We recruited 600 participants using \hl{Prolific.co}\cite{palan2018prolific}, following the same recruitment and exclusion criteria as for Experiment 1. However, for Experiment 2B, since there was a risk for participants' political affiliations to affect their priors or interpretation of the presented stimuli, we balanced the participant pool by ensuring that both sides of the political spectrum in the US - Democrats and Republicans - were equally represented.  After excluding participants who had failed attention checks or had missing data or nonsensical response, we ended up with 294 participants ($\mu_{age} = 42.06$, $\sigma_{age} = 14.69$, \text{149 women}) for Experiment 2A, and 298 participants ($\mu_{age} = 42.60$, $\sigma_{age} = 14.53$, \text{145 women}) for Experiment 2B. The participant pool also had a good variance in education levels (see SM). \hl{Participants were compensated at a flat rate of \$2.54 for Experiment 2A  with an expected completion time of 17 minutes and at \$2.38 for Experiment 2B  with an expected completion time of 15 minutes. These rates were decided to keep in line with the average hourly wage of about \$9.52 at Prolific.co prevalent at the time of the survey. The difference in the compensation rates for the two Experiments was due to the updating of the expected time needed for Experiment 2B based on the time taken by participants in Experiment 2A. However, since our time expectations were slightly generous, the participants were practically compensated at \$9.80 per hour for their effort.}

\subsection{Design and Materials}
We presented participants with two line charts. These charts were based on a hypothetical scenario involving data from publications from two equally reputable institutes on the relationship between crop yield and water use for Experiment 2A and between the quality of life and public spending for Experiment 2B. As with Experiment 1, the presented charts differed in their strength and direction of the relationship between the two variables. We included the same six Conditions and counterbalancing methods as used in Experiment 1.

We created two sets of line charts for Experiments 2A and 2B. By keeping the underlying simulated distribution for the charts the same, we changed the axes scales between Experiment 2A and 2B to make the scenarios more realistic. For Experiment 2A, scale values were based on research articles on water use efficiency in agriculture \cite{sharma2015water}\cite{de2011increasing}\cite{hatfield2019water}, and actual public expenditures for counties in the United States were used for scales in Experiment 2B \cite{public2019spending}. For Experiment 2A, crop yield was expressed on a scale of 0 to 100 lbs per acre, and water use was expressed on a scale of 0 to 100 inches. For Experiment 2B, the axes represented the quality of life, measured on a scale of 0 to 100 points on a hypothetical `Quality of Life' index, and public spending expressed on a scale of 0 to 1000 million USD.  

\subsection{Procedure}
Similar to Experiment 1, participants were presented with a short introduction to the scenario and instructions to use the custom-built widget to report their synthesized slopes. We also used this widget to capture their \textit{prior beliefs} on the relationship between crop yield and water use for an unknown plant for 2A and between the quality of life and public spending for 2B. The rest of the Experiment was similar to Experiment 1, with the stimuli and questions modified to reflect changes in the chart information.

\subsection{Results}

We analyzed participants' divergence from the mathematical average of the two stimuli. As shown in Figure \ref{fig:mathematicalAverage_exp1}, for Conditions \textcolor{highNegative}{2}, \textcolor{bothStrong}{3}, and \textcolor{highNegative}{6}, including a \textcolor{highNegative}{High negative} stimulus, we found a statistically significant divergence from the mathematical average with a small to moderate effect size for both Experiments 2A and 2B. Additionally, for Experiment 2B, we found one more Condition (\textcolor{highPositive}{Low negative/High positive}) that had a statistically significant divergence from the mathematical average with a small effect (Conditions \textcolor{highPositive}{1}, \textcolor{highPositive}{4}, and \textcolor{highPositive}{5} for $x=20$ and $x=80$).

\vspace{-0.5mm}
\subsubsection{Triangulation} 
Following Experiment 1 for triangulation, we computed the values from the widget for the specific points ($x = 20$ and 80, and $y = 50$) that we asked for using the text input from participants and compared them to see if there were statistically significant differences. We found statistically significant differences by running a paired sample t-test on the three abscissa values. Also, to note, we had to discard the data collected by one of the text boxes (for $y = 50$) for Experiment 2B due to a technical error in logging the data. To investigate if these differences were localized within some Conditions, we ran 12 independent t-tests for both abscissas \hl{- $x = 20$ and $x = 80$. We also ran six other t-tests for each value, accounting for order effects.} The differences were found to be localized to some specific Conditions (Experiment 2A: Conditions \textcolor{highPositive}{1}, \textcolor{highPositive}{4} and \textcolor{highPositive}{5} for $x=20$, $x=80$; Experiment 2B:  Condition \textcolor{highNegative}{2} for $x=20$, and Conditions \textcolor{highPositive}{1}, \textcolor{highPositive}{4} and \textcolor{highPositive}{5} for $x=80$).

In a separate analysis, we computed the slope from the values that participants input using text boxes (\hl{Synthesis Q1}) and compared it against slope values drawn using the widget (\hl{Synthesis Q4}). For 5 out of the 12 Conditions, the slopes drawn using the widget and those computed using the \hl{text boxes} were statistically significantly different (p < 0.05; see SM for details).

As with Experiment 1, we triangulated visual information synthesis using a decision-making task (\hl{Synthesis Q3}). For Experiments 2A and 2B, all categories had over $60\%$ of agreement, with the highest agreement between widget and text boxes (2A: $95\%$; 2B: $98\%$). All other categories had $62-77\%$ of agreement levels. Thus, the different modes of capturing participants' mental models were functioning equally well, with a stronger relationship between data captured through text box and widget. 

\vspace{-0.5mm}
\subsubsection{Priors} 
Most people reported priors with a mildly positive relationship between the two variables (Exp 2A:$\mu_{prior} = 0.31,\sigma_{prior} = 0.28, SE_{prior}= 0.017$, Exp 2B: $\mu_{prior} = 0.35,\sigma_{prior} = 0.27, SE_{prior}= 0.016$.) We performed a one-way ANOVA on the prior slope response, but found no significant differences among priors across all Conditions (2A: $Pr(>F) = 0.752$, 2B: $Pr(>F) = 0.884$). However, we found statistically significant differences between participants when considering their political beliefs for Experiment 2B, with Republicans reporting priors depicting a weaker relationship between public spending and quality of life ($\mu_{R} = 0.321$, $\mu_{D} = 0.393$; $Pr(>F) = 0.0224$).

\vspace{-0.5mm}
\subsubsection{Order}
We computed the normalized distance from the stronger slope (larger absolute value) and measured the effects of the order in which the stronger slope was presented on this distance metric through a linear regression model. We did not find any effects of order for Experiments 2A and 2B (DV: Normalized distance from stronger slope, IV: Order of the second slope; Experiment 2A - $R^2_{adj} = 0.001$, $p = 0.2442$; Experiment 2B - $R^2_{adj} = -0.0005$, $p = 0.3595$)

\vspace{-0.5mm}
\subsubsection{Direction of the presented stimuli}
As shown in Figure \ref{fig:effectDirection}, when the two slopes were in the same direction (both positive or both negative), the synthesized slope weighted the stronger slope (2A: $\mu = 38.52\%$, 2B: $\mu = 38.89\%$) and the weaker slope (2A: $\mu = 61.47\%$, $SE = 3.04\%$,; 2B: $\mu = 61.11\%$, $SE= 3.08\%$ ), accounting for order effects, when analysed together. However, these weights increased when both slopes were negative, with even more weight assigned to the smaller slope (2A: $\mu = 86.62\%$, $SE = 9.26\%$; 2B: $\mu = 104.53 \%$, $SE = 10.40\%$). When the two slopes were in the opposite direction, the synthesized slope weighted the positive slope (2A: $\mu = 59.73\%$, $SE = 2.61\%$; 2B: $\mu = 68.66\%$, $SE = 2.60\%$), accounting for order effects.

\vspace{-0.5mm}
\subsubsection{Discrepancy between the presented stimuli}
Similar to Experiment 1, we observed a \textcolor{highNegative}{High negative} - \textcolor{highPositive}{High positive} divide across all three distance metrics (see Figure \ref{fig:errorBars}). Across Experiments, Conditions with a \textcolor{highPositive}{High positive} slope were \hl{on the left side} of Conditions with a \textcolor{highNegative}{High negative} slope. As in Experiment 1, there was an increasing trend for distance from the mathematical average, i.e., as the difference between stimuli slopes increased, the response slopes tended to move away from the mathematical average. However, an opposite effect occurred for the stronger slope. As the difference between stimuli slopes increased, the response slope tended to move closer to the stronger slope.
\section{Experiment 3 --- Synthesizing Text and Visuals}
We investigated interactions between visualization and text on synthesis behaviors. This Experiment simulates real-world scenarios to afford increased ecological validity. 

\vspace{-0.5mm}

\subsection{Participants}
We recruited 288 participants using \hl{Prolific.co}\cite{palan2018prolific}, following the same criteria as for Experiment 2B, keeping an equal representation of different sides of the political spectrum. After excluding participants who had failed attention checks or had missing data or nonsensical responses, we ended up with 279 participants($\mu_{age} = 38.7$, $\sigma_{age} = 13.2$, \text{138 women}). The participant pool also had a good variance in education levels (see SM). \hl{Participants were compensated at a flat rate of \$2.70 with an expected completion time of 17 minutes, keeping in line with the average hourly wage of about \$9.52 at Prolific.co prevalent at the time of the survey. However, since our time expectations were slightly generous, the participants were practically compensated at about \$11.60 per hour for their effort.}

\vspace{-0.5mm}

\subsection{Design and Materials}
We presented participants with two news articles about the relationship between quality of life and public spending, with accompanying charts. The charts were the same as the ones used in Experiment 2B. 
We included the same six Conditions and the same counterbalancing methods as before. 
The authors wrote the articles to precisely control text characteristics, provide only necessary variation to mirror the strength of the relationships shown in the charts, and give the articles a realistic feeling while keeping all other aspects constant. We also needed articles that participants had not read before. Within Conditions, we also varied the design of the two articles to make participants believe that the information was taken from two different news websites. However, to keep the design of the Experiment simple, we could not counterbalance in terms of article design (e.g., colors, news outlet), such that the first article always had the same design, and the second article always had the same design. The articles underwent multiple iterations of scrutiny and input from all four authors and two undergraduate research assistants to ensure that all the constraints and conditions were met.  We have included the articles in the SM. 

\vspace{-0.5mm}

\subsection{Procedure}
We followed a similar procedure as before, with slight changes in questions and stimuli presented. Participants started the Experiment with informed consent, followed by a screen resizing widget to adjust the resolution. They were then presented with a short introduction to the scenario and instructions to use the custom-built widget described previously. The participants were then presented with the articles one by one. As before, in the prompt, we emphasized attending to the information presented in the chart and building a mental model for the relationship between the variables. However, unlike in previous Experiments, for Experiment 3, participants were not asked to read the chart and compute the abscissa for a given ordinate. The rest of the data collection occurred similarly to Experiment 2B.

\vspace{-0.5mm}
\subsection{Results}

As for prior Experiments, we captured participants' information synthesis using a custom-made widget. Participants were asked to report their estimation of the relationship between the two variables of interest before and after the presentation of the stimuli. A graphical representation of their responses is in SM. As in previous Experiments, for Experiment 3, each Condition was counterbalanced based on the order in which the stimuli were presented. 

We analyzed participants' divergence from the mathematical average of the two stimuli. For Conditions \textcolor{highPositive}{1}, \textcolor{highNegative}{2}, \textcolor{bothStrong}{3}, and \textcolor{highNegative}{6}, we found a statistically significant divergence from the mathematical average with a large effect size for Conditions \textcolor{highPositive}{1} and \textcolor{highNegative}{6} (\textcolor{highPositive}{High positive/ Medium positive}, and \textcolor{highNegative}{High negative/Low positive}) and a small effect for others (see SM for details). 

\vspace{-0.5mm}

\subsubsection{Triangulation}
We followed a similar approach as the previous Experiments for triangulating, first by computing the values from the widget for the specific points ($x = 20$ and $80$, and  $y = 50$) that we asked for using the text input from participants and comparing them to see if there were statistically significant differences. We found statistically significant differences by running a paired sample t-test on these three different coordinates ($x = 20$ and $80$, and  $y = 50$). To investigate if these difference were localized within some Conditions, we ran 12 independent t-tests for \hl{each of the three values - $x = 20$, $x =80$, and $y=50$. We also ran six other t-tests for each value, accounting for order effects}. The differences were found to be localized to High positive and Medium positive and their negative counterparts, and all Conditions for $y = 50$.

In another analysis, we computed the slope from the values that participants inputted using text boxes (\hl{Synthesis Q1}) and compared it against the slope value drawn using the widget(\hl{Synthesis Q4}). We found for 4 out of the 12 Conditions, the slopes drawn using the widget and the ones computed using the abscissa values were statistically significantly different(p < 0.05; see SM for details).

As with prior Experiments, we triangulated visual information synthesis using a decision-making task(\hl{Synthesis Q3}). For Experiment 3, all categories had over 68\% agreement, with the highest agreement between widget and text (98\%). All other categories had 68-79\% of agreement levels. 

\vspace{-0.5mm}

\subsubsection{Priors}
Most people reported priors with a mildly positive relationship between the two variables ($\mu_{prior} = 0.37, \sigma_{prior} = 0.28, SE_{prior}= 0.017$). We performed a one-way ANOVA on the prior slope response and found no statistically significant differences between Conditions ((Pr>F) = 0.239). However, we found statistically significant differences between participants when considering their political beliefs, with Republicans reporting to have priors depicting a weaker relationship between public spending and quality of life ($\mu_{R} = 0.291$, $\mu_{D} = 0.421$; $Pr(>F) = 4.14e-07$).

\vspace{-0.5mm}

\subsubsection{Order}
We computed the normalized distance from the stronger slope and measured the effects of the order in which the stronger slope was presented on this distance metric through a linear regression model.(DV: Normalized distance from stronger slope, IV: Order of the second slope $R^2_{adj} = 0.0004$, $p = 0.29$). We did not find any order effects.

\vspace{-0.5mm}
\subsubsection{Direction of the presented stimuli}

As shown in Figure \ref{fig:effectDirection}, when the two slopes were in the \textit{same} direction (both positive or both negative), the synthesized slope weighted the weaker slope $61.78\%$ ($SE = 2.89\%$), accounting for order effects, when analyzed together. However, \hl{these differences in weighting became even more pronounced} when both slopes were negative, with even more weight assigned to the weaker slope ($\mu = 85.29\%$, $SE = 11.83\%$).
When the two slopes were in the \textit{opposite} direction, the synthesized slope weighted the positive slope $68.25\%$($SE = 2.91\%$), accounting for order effects. These results are in line with the results observed for Experiment 2B.

\vspace{-0.5mm}
\subsubsection{Discrepancy between the presented stimuli}

Results for this section were consistent with our observations for Experiments 1, 2A, and 2B. We observed a \textcolor{highNegative}{High Negative} - \textcolor{highPositive}{High Positive} divide across all three distance metrics (see Figure \ref{fig:errorBars}). Across Experiments, Conditions with a \textcolor{highPositive}{High positive} slope were \hl{on the left side} of Conditions where a stimulus had a \textcolor{highNegative}{High negative} slope. Similar to prior Experiments, there was an increasing trend to distance from the mathematical average, i.e., as the difference between stimuli slopes increased, response slopes tended to move away from the mathematical average. However, an opposite effect occurred for the stronger slope, i.e., as the difference between stimuli slopes increased, the response slope tended to move closer to the stronger slope.
\begin{figure}[t!]
\setlength{\textfloatsep}{4pt plus 1.0pt minus 1.0pt}
\setlength{\intextsep}{4pt plus 1.0pt minus 1.0pt}
\setlength{\floatsep}{4pt plus 1.0pt minus 1.0pt}
\centering
  \includegraphics[width=\textwidth]{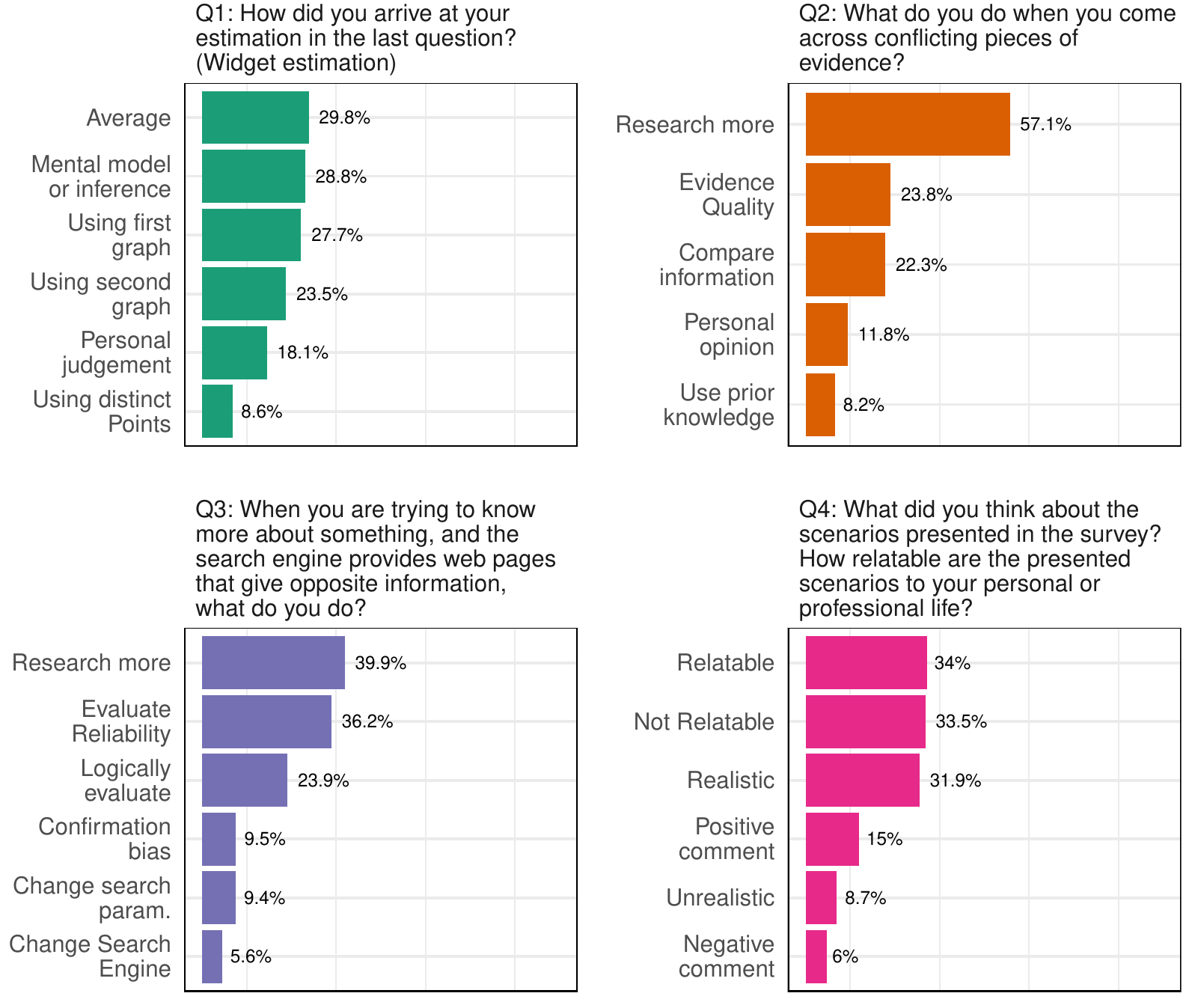}
  \caption{Quantitative summary of participants' responses (\hl{N = 1166}) on 4 qualitative questions, categorizing their strategies for tackling conflicting information, and their thoughts on the Experiments 1-3}
    \label{fig:exp4Coded}
\end{figure}

\section{Analysis of Real-World Synthesis Behaviors}
\label{Exp4QualData}
We analyzed the various strategies participants reported for synthesizing conflicting information in their everyday life and their opinions about the relatability of Experiments 1, 2A, 2B, and 3. 
\vspace{-2mm}
\subsection{Participants}
The same participants (\hl{N = 1166}) recruited for Experiments 1, 2A, 2B, and 3 responded to a set of qualitative questions asking how they synthesize conflicting information and the strategies they use, both for our Experiments and in real life. 

\vspace{-0.5mm}
\subsection{Questions and Analysis Approach}
We asked four qualitative questions in each of Experiments 1-3, with a response requirement of a minimum of 20 characters.
The specific questions are shown in Figure \ref{fig:exp4Coded}.

We used an inductive thematic analysis for coding the collected data\cite{braun2006using, braun2012thematic} to preserve the richness of information while making the analysis cognitively manageable. 
We recruited two undergraduate research assistants to independently construct a set of codes for each question based on a set of pilot data ($N = 78$). 
Next, they conducted a converging exercise to integrate their independent codes into a standard set of codes, with definitions and prototypical examples. They then used these codes to independently code the participant responses for all four questions.

\vspace{-0.5mm}
\subsection{Results: Strategies for Slope Estimation}
We asked participants about their strategy for their estimations in the slope synthesis task (via widget) in Experiments 1-3. 
The main strategies we identified were (a) averaging both charts, using either (b) the first or (c) the second chart only, (d) inferring from the charts using a mental model, (e) using distinct points or values the participant remembered, and (f) using personal judgment or intuition. 
There were also `Irrelevant' strategies for categorizing responses that were nonsensical, blank, hinted toward the participants being confused as to what the question was asking, or reported having technology issues. 
Figure \ref{fig:exp4Coded} provides a quantitative summary. 
Participants were able to report using multiple strategies. Thus, the percentages in the figure can add up to over $100\%$. 

Almost $30\%$ of participants reported using an average or some other mental model for arriving at their estimates. As an example to underscore this, one participant from Experiment 1 wrote, \textit{``Both of the charts showed a decline of Y as X increased, however the sharpness of the drop varied between the two charts. I tried my best to show a middle ground based on my memory of the charts.''} 

Many participants based their estimation on favoring one chart/article or the other. For example, one participant from Experiment 3 wrote, \textit{``I think the first article is more true and factual than the second. The second seems to be a very one-sided political argument.''} 
Some participants used distinctive points for arriving at their estimates. One participant from Experiment 2A mentioned, \textit{``I tried to average the two graphs shown previously. One started at 20 the other at 60, so I made the start point at 40.  The end point was about 85 for one and about 20 for the other, so I made 52 my end point.''} 
Some also relied on their intuition. For example, one participant from Experiment 2B wrote, \textit{``It was hard for me to believe Chart A, that the more is spent, the lower the quality of life, so I discounted it.  Chart B was more believable, so I lowered my expectations and lowered my original chart by a bit, to semi-match Chart B.''}

\vspace{-0.5mm}
\subsection{Results: Conflicting Evidence}
Participants were asked about their strategies to cope with conflicting evidence. Based on pilot data, we identified 5 categories - (a) researching more, (b) using prior beliefs or personal opinion to come to conclusion, (c) using prior knowledge and/or personal experiences, (d) looking at the quality of evidence and the credibility and reliability of the sources,  (e) comparing information and looking for similarities, differences and reasoning used to justify the evidence. As before, we also had a category for irrelevant responses (Figure~\ref{fig:exp4Coded}).

A majority of participants reported researching for more information when faced with conflicting evidence, mentioning that \textit{``I will research the topic more to figure out which side is correct''}, \textit{``I try to read more about the topic to gather my own opinion based on several sources.''}, or \textit{``I continue doing more research until I can find what the truth is, or at least which side is more favored.''}

A substantial proportion also reported using evidence quality as a metric to arrive at the truth. One participant from Experiment 3 wrote, \textit{``I try to evaluate which source of information is the most trustworthy, and then try to see if any other reliable sources have recreated a certain piece of evidence. I want to know about the veracity of each side before deciding which to believe''}. 

Participants also talked about comparing information. One participant from Experiment 2A responded, \textit{``I consider each separately and try to see if there is a connection or some part that corresponds to each other, then look at the differences.''} A few participants reported using their personal beliefs, opinions, or prior knowledge as a valid method for resolving this conflict. Typical responses under this category include, \textit{``I use my best judgment and go with my gut instinct.''}, \textit{``I use my common sense to determine which is more accurate.''}, or \textit{``Use my experience and feel into all aspect of what I know with my intuition.''}

\vspace{-0.5mm}
\subsection{Results: Conflicting Information In Online Searches}
Participants reported what they do when online searches yield conflicting information. We identified 6 strategies - (a) evaluating the reliability and reputability of the website or sources, (b) using a different search engine, (c) changing search parameters, (d) using logical or scientific reasoning to evaluate the veracity of the claims, (e) researching more, and (f) using information that matches their prior beliefs.

As shown in Figure \ref{fig:exp4Coded}, researching and looking for reliability/quality of information were the most popular strategies for resolving conflicting information online.
One participant from Experiment 2B described their strategy as, \textit{``I learn what the opposite information is, and try and see if I think it is valid''}. There was also a noticeable proportion of participants who were cognisant of their confirmation bias, and some even embraced it. One participant from Experiment 2B wrote, \textit{``I usually try to find information that would validate my beliefs.''} 

Some participants reported changing the search parameters or the search engine to access better information. One participant from Experiment 1 wrote, \textit{``If they provide information that isn't relevant to my search query I will refine the query until I get higher quality information from more credible sources.''}

\vspace{-0.5mm}
\subsection{Results: General Feedback}
To gather feedback on Experiments 1-3, we asked participants about their views on the relatability of the scenarios in the Experiments and derived the following categories - (a) Relatable, (b) Not relatable, (c) Plausible or realistic, (d) Unrealistic, (e) Positive feedback or comments, and (f) Negative feedback or comments. 

As shown in Figure \ref{fig:exp4Coded}, roughly 1/3 of participants reported finding the scenarios relatable. A participant from Experiment 3 wrote, \textit{``I think they're very relatable, as I live in a larger city than I did when I was growing up, and I can definitely tell that the more public spending that is done in certain areas definitely affects the quality of life of the people living there.''}. 

Some participants saw the scenario as unrelatable. Typical responses in this category include, \textit{``Not relatable to me. The only experience I have with plans is watering house plants.''}, \textit{``I had no issues with the scenarios. I don't think the scenarios relate too much to my personal life.''} (from a participant from Experiment 2B), and \textit{``These aren't applicable to my personal or profession life''} (Experiment 3). 

Some participants commented on the realistic nature of the scenarios. One participant from Experiment 3 wrote, \textit{``I thought these scenarios were pretty realistic, as each article didn't necessarily adhere to any bias. I didn't get a sense of "oh, this article is definitely written by Republicans" and vice-versa; it seemed like they were neutral and unbiased, and both generally cared about the well-being of the people.''}. 
On the other hand, a small number of participants felt the scenarios were unrealistic. 
For example, one participant from Experiment 2B wrote, \textit{``Contradicted my beliefs but I went with it for the sake of the survey. Not sure if the data is accurate.''}. Another from Experiment 2A wrote, \textit{``I think the scenarios were slightly unrealistic and they were not at all related to my personal or professional life.''}


\section{Limitations and Future Work}
\label{limitationsFuture}

As an initial examination of how readers synthesize conflicting information, it is too early to provide concrete design guidelines to mitigate bias. We discuss several limitations in our study and provide promising future research directions.
\newline
\vspace{-2mm}

\noindent \textbf{Generalizable experimental Designs:} To mimic real-world scenarios for synthesizing conflicting information, our experiments covered a range of topics and formats, including adding accompanying text to visualizations to simulate news articles.
We did not vary the design of the article layout besides a minor color change, but in the real world, sources often look more dissimilar. 
Additionally, we controlled for the type of visualization used by asking participants to read similar line charts with overlaid scattered dots. 
These design choices limit our ability to generalize our findings to other visualization types, as visualization techniques can profoundly impact human perception and decision-making \cite{xiong2021perceptual, xiong2019illusion, xiong2021visual}. 
Furthermore, we did not vary the number of data points in the current experiments and only asked participants to synthesize two pieces of information. In contrast, in the real world, people are expected to interpret multiple pieces of information using datasets with varying sample sizes.
We also used fictional scenarios that maximized control over realism. 
Future work can additionally investigate the effects of the aesthetic design of information sources, different visualization techniques, sample sizes, visualization topics, and the number of visualizations on reader behavior to create a more comprehensive model of information synthesis.
\newline
\vspace{-2mm}

\noindent \textbf{Slope vs. Intercept Analysis:} In the present study, we considered multiple measures of capturing participants' mental models for synthesized data, such as asking participants to make predictions or compare data trends based on their synthesized data. 
We also asked them to draw a synthesized slope and closely examined how the drawn line slope differed. 
We recognize that slopes might not be the only information from the visualizations participants can synthesize. 
For example, participants might weigh the line intercept differently depending on the strength of the slope. 
They might be influenced by the scattered dots plotted in the background or rely on the angle between the line and the x-axis. 
While we collected intercept data, because our experimental conditions were set up based on slope categories, the intercept data became less relevant to our research question. 
Nevertheless, we have made the intercept data collected available in our SM for scholars interested in further analysis. 
Future work can manipulate intercepts instead of slopes in this type of experiment to build on our current model of information synthesis.
Another way to look at the data could have been through the lens of the angle between lines. We recognize that the slope adds a non-linear trigonometric function. 
However, for our experiments, the stimuli slopes ranged between $-32^\circ$ to $+33^\circ$, and we did not have to deal with the asymptotic behavior of the tangent function at higher angles. 
Furthermore, there still exists a knowledge gap in understanding the perception pathways in recognizing the angle between the x-axis and the line vis-\`a-vis the strength of the relationship shown through the slope parameter.
\newline
\vspace{-2mm}

\noindent \textbf{Effect of Text and Metadata:} We added text to accompany visualization in experiment 3 to simulate real-world scenarios. 
We did not consider the effect of only showing textual information nor the effect of interaction between text and visualizations on synthesis behavior, as they would be a separate set of research questions. 
For example, depending on the conditions participants were assigned (e.g., high positive slope or low negative slope), the accompanying text varied to fit the patterns in the visualization. 
From existing work, we know that information consumers often synthesize text and visual information to enhance understanding \cite{carney2002pictorial, mayer2002multimedia}, and that text and visualizations do interact to impact reader interpretation (e.g., \cite{kong2018frames, kim2021towards}). 
Our work serves as an initial step in the cross-section between visualization and text information synthesis, and future work should further explore the interplay between visuals and text.  
\newline
\vspace{-2mm}

\noindent \textbf{Response Modality and Feedback:} We are interested in how participants organically behave when synthesizing information. Thus we did not especially incentivize participants to provide an accurate synthesis of the two charts.
Participants also indicated the most probable slope value and were not prompted to provide a range of possible slope values.
Future work can explore whether providing incentives or enabling participants to indicate some uncertainty intervals can mitigate the systematic bias we observed. Recent work on using Bayesian cognitive frameworks to capture uncertainty may be beneficial \cite{kim2019bayesian}.
Additionally, providing feedback on their response through several practice trials may help participants realize their biases and help them synthesize information more fairly.

Participants were given a consent form informing them about the study's objectives, the associated risks and benefits, and an option to withdraw from the experiment at any point. 
They were also encouraged to send us messages on the Prolific platform should they have questions or concerns. 
However, an explicit debriefing at the end of the survey was not provided since the study involved hypothetical scenarios, which is common in these types of behavioral studies. 
We constructed these scenarios in a manner to minimize the risk of participants developing strong opinions or being harmed based on the information provided through the experiment. All the experimental stimuli were constructed and presented in a manner that did not provide specific information as a 'fact of the matter.' For example, we described the two institutions providing the data to synthesize as institutes A and B, and the plant name in the water usage scenario was made up. 
However, in Experiment 3, our hypothetical scenario about life quality included reference to real entities (counties in the United States). It is possible that participants may be influenced by this information.
Future studies should explicitly debrief participants on all hypothetical scenarios presented and reflect on the potential impact of showing participants false or contrived information. 
\vspace{-1mm}
\section{Conclusion}
\label{conclusion}
When synthesizing information across two line charts depicting conflicting information, participants considered both the direction and the slope of the lines. When both slopes are in the \textit{same} direction, people tend to more heavily weigh the \textit{weaker} relationship. 
When slopes are in \textit{opposite} directions, participants tend to favor the \textit{positive} relationship. This effect tends to increase with additional informational and contextual clues presented along with the charts. However, as the discrepancy between stimuli slopes increases, participants tend to favor the stronger slope more. Overall, the general synthesis patterns remained consistent across our experiments, suggesting strong underlying mechanisms behind visual synthesis with line charts. 
\vspace{-2mm}
\section{Acknowledgments}
\label{acknowledement}
We thank our reviewers for their helpful feedback. We also thank Gabriella Nugent and Lynn Li for assisting with stimuli for experiment 3, and Kylie Lin, for their assistance with coding participants' responses to the qualitative questions. 


\bibliographystyle{abbrv}
\bibliography{99_refs}

\balance{}










\end{document}